\documentstyle[prl,aps]{revtex}
\begin{document}
\draft
\title{
The charge radii and the decay rates of the pseudoscalar meson}
\author{Yongkyu Ko \footnote{Electronic address: yongkyu@phya.yonsei.ac.kr}}
\address{
Department of Physics, Yonsei University, Seoul 120-749, Korea}
\date{\today}
\maketitle
\begin{abstract}
The charge radii and the decay rates of the pion and kaons are calculated,
using the relativistic equation of motion with a linear potential.  Those 
physical quantities are quite well explained with the current quark masses
in the case of the pion.  It is found that the Van Royen-Weisskopf paradox 
can be cleared out only in the linear potential model by considering the 
color degree of freedom of the quark in the meson.  The physical quantities 
for the kaon are not so satisfactory that it should be required to reconsider 
the Cabibbo-Kobayashi-Maskawa matrix element in the kaon decay.
\end{abstract}
\vspace{1cm}
Recently the author has attempted to derive a quark confinement potential
from quantum chromodynamics (QCD) \cite{Ko}.  
Unfortunately the attempt does not yet succeed, because
there is a serious problem to define the Fourier transform of a confinement
potential which is non-local.  However it is natural to think that there should
exist a certain confinement potential in nature, considering hadronic phenomena.
So there are some attempts to define it by introducing a small parameter
\cite{Hershach,Eyre}.  This letter has the purpose to explain some physical
quantities using the potential in order to confirm 
the existence of such a confinement potential in nature.
The electromagnetic charge radii and the decay rates of the pseudoscalar meson
are very important physical quantities for understanding
the structure of the hadron and QCD of its theory.  Those physical quantities
have been calculated many times since the advent of the quark model 
\cite{Weiss,Kokkedee,Lucha,Allen,Krutov}.
However there are some difficulties to describe hadronic phenomena,
such as the constituent quark masses and the Van Royen-Weisskopf paradox 
in those literatures.
This letter shows that the proper relativistic description explains 
such difficulties quite reasonably with the current quark masses.  The 
consideration of the color degree of freedom clears out the paradox excellently, 
which is the difference of
the spatial wave function $|\Psi(0)|^2$ between the calculations of the charge 
radius and the decay rate of the pion.  In the case of the kaon, however, 
there is a difficulty to explain its charge radii and decay rate 
consistently.  The reason is conjectured that the Cabibbo-Kobayashi-Maskawa 
(CKM) matrix element $|V_{us}|^2$
is too large so that the kaon decay constant is too small.  Therefore it is
necessary to scrutinize the CKM matrix element for the kaon decay.

In order to describe the relativistic feature of the quark  in the bound
system of the meson, the Klein Gordon equation is used to do it rather than
the Dirac equation.  Since the quark is, of course, a fermion, the Dirac
equation should be used to do it.  However there are good reasons to adopt
the Klein Gordon equation, because the degeneracy of spin singlet and 
triplet states is broken larger than the masses of pseudoscalar mesons so 
that the spin structure of the equation of motion may be less important.  
Moreover the 
relative motion of the quark in the meson can be assumed to be governed by the 
effective potential of the spin singlet state.  While the electron in the 
hydrogen atom has two spin states to occupy in the ground state regardless of 
the spin of the proton, the quark has only one spin state in the ground state
of the meson due to the influence of the other quark,
because they have similar masses.  This is another reason to adopt the Klein
Gordon equation, because it is not necessary to describe the motion of the 
quark with a spinor.

The Klein Gordon equation for a free quark is given by
\begin{equation}
(\Box + m^2) \Psi =0.
\end{equation}
Since two body system can be separated into the relative and the center of mass
parts, if the equation is modified as follows
\begin{eqnarray}
& &-{1 \over m_1} \nabla_1^2 \Psi-({E_1^2 \over m_1}-m_1) \Psi=0 \nonumber \\
& &-{1 \over m_2} \nabla_2^2 \Psi-({E_2^2 \over m_2}-m_2) \Psi=0.
\end{eqnarray}
Using the identity
\begin{equation}
{1 \over m_1} \nabla_1^2+{1 \over m_2} \nabla_2^2
={1 \over{\mu}} \nabla_r^2+{1 \over{m_1+ m_2}} \nabla_R^2,
\end{equation}
where $\mu$ is the reduce mass of $m_1$ and $m_2$, the sum of the two equations
is rewritten as
\begin{equation}
-{1 \over{\mu}} \nabla_r^2 \Psi - {1 \over{m_1+ m_2}} \nabla_R^2 \Psi
-({\mbox{\boldmath{$p$}}_1^2 \over m_1}+{\mbox{\boldmath{$p$}}_2^2 \over m_2}) 
\Psi=0.
\end{equation}
At the center of mass frame, $\nabla_R^2 \Psi=0$ and $\mbox{\boldmath
{$p$}}_1^2=\mbox{\boldmath{$p$}}_2^2=\mbox{\boldmath{$p$}}^2$, 
the equation is written simply as
\begin{equation}
-{1 \over{\mu}} \nabla_r^2 \Psi 
-{\mbox{\boldmath{$p$}}^2 \over{\mu}} \Psi=0,
\end{equation}
where $\mbox{\boldmath{$p$}}^2=E_1^2-m_1^2=E_2^2-m_2^2=E_r^2-\mu^2
=\mbox{\boldmath{$p$}}_r^2$.  
The relative energy-momentum and mass are 
$E_r=m_2E'_1/(m_1+m_2),
\mbox{\boldmath{$p$}}_r=m_2\mbox{\boldmath{$p$}}'_1/(m_1+m_2)$, 
and $\mu=m_2m_1/(m_1+m_2)$, where 
$E'_1$ and $\mbox{\boldmath{$p$}}'_1$ are the energy and
momentum in the rest frame of $m_2$. 
Thus the relative part of the 
two body system is reduced to the equation of motion of a single particle:
\begin{equation}
(\Box + \mu^2) \Psi =0,
\end{equation}
which satisfies the relative energy-momentum relation.
Since the two particles interact with each other with exchanging their gauge
field, the derivative should be replaced to the covariant derivative as follows
\begin{equation}
(D_{\mu}D^{\mu} + \mu^2) \Psi =0.
\end{equation}
where $D_{\mu}=\partial_{\mu}-igA_{\mu}$.  A few algebraic effort can make the
equation be decomposed as
\begin{equation}
-{1 \over{E_r+\mu}} \nabla^2 \Psi+{1 \over{E_r +\mu}}(2 gA_0E_r -2g 
\mbox{\boldmath{$A$}} \cdot \mbox{\boldmath{$p$}}_r -g^2 A_0^2 +g^2 
\mbox{\boldmath{$A$}}^2) \Psi =(E_r -\mu) \Psi, \label{releq}
\end{equation}
where the color index is hidden in the case of QCD.
For the test of the validity of the equation, let's approximate it for the 
case of QED $E \approx \mu$, namely, $E_r=\mu+K_r$ and $K_r\ll \mu$.  
Thus the equation is expanded
as
\begin{equation}
-{1 \over{2 \mu}} \nabla^2 \Psi+ eA_0 \Psi- {e \over{\mu}}\mbox{\boldmath{$A$}} 
\cdot \mbox{\boldmath{$p$}}_r \Psi +{e^2 \over{2\mu}} \mbox{\boldmath{$A$}}^2 
\Psi -{1\over{2\mu}}[(eA_0)^2-(K_r eA_0)+\{K_r(-{\nabla^2\over{2\mu}})\}]\Psi
=K_r \Psi,
\end{equation}
where higher order terms are ignored.  If the potential $A_0=-e /r$ is 
inserted, this equation is nothing but the Schr\"odiger equation with the 
relativistic correction terms in the square bracket which agree to that
in the standard texts by inserting the relation $K_r =-\nabla^2/2\mu+eA_0$ 
in the brace bracket.  In the nonrelativistic 
case of the motion of the particle, the potential which make a bound system 
is the coulomb one.  However the potential should be modified as the 
effective potential as insisted in Ref \cite{Ko} due to the effect of the
vacuum polarization of the quark, when the motion of the
particle reaches to the ultra relativistic region.  Hence Eq. (\ref{releq}) 
can be written as
\begin{equation}
- \nabla^2 \Psi+2 E_r(V_{\text{eff.}} -E_r^2+\mu^2) \Psi =0.
\end{equation}
Notice the coefficient $2E_r$ in front of the effective potential which is 
$2 \mu$ in the
nonrelativistic case.  It seems to be the origin to introduce the constituent
quark masses in the calculations of low energy physical quantities with
the Schr\"odinger equation and other relativistic approaches.  
For the confinement potential $V_{\text{eff.}}=kr$, where
$k=2 \alpha_s^2/3\pi \Sigma_f m_f^2$ under the condition 
$4m_f^2<q_{\text{gluon}}^2$
in Ref. \cite{Ko}, the
solution of the differential equation is just the Airy function which is
\begin{equation}
\Psi(r)=C f(r) e^{-ar^{3/2}},\label{wave}
\end{equation}
where $C=\sqrt{3a^2/2\pi}$ for $f(r)=1$ which gives the finite value 
$|\Psi(0)|^2$ in the calculations of decay rates.  The parameter $a$ is 
calculated as $\sqrt{8E_rk/9}$ from the differential equation
for the ground state of the quark.  Since the coefficient of the wave function
depends on the energy of the particle instead of its mass, 
it is natural that the minima for the
s- and d-wave components of the deuteron function should be shifted toward
large momenta in relativistic case as pointed out in Ref. \cite{Kamada}.  In
this reference, the potential should be also transformed or rewritten by a scale
transformation in the momenta, but such a strange thing does not occur in this
approach, because the coefficient $2E_r$ plays the role sufficiently.  Moreover
no Lorentz contraction leads to no transformation, because the variable $r$ of 
the potential is perpendicular to the motion of the particle.

The form factor and the charge radius are defined and calculated for the
wavefunction as 
\begin{eqnarray}
F(\mbox{\boldmath{$q$}}^2) &=& \int \rho(r) e^{i \mbox{\boldmath{$q$}}
\cdot \mbox{\boldmath{$r$}}} d^3r 
= Q - {1\over6}q^2 <r^2> + \cdot \cdot \cdot \cdot \cdot \cdot,\nonumber\\
 <r^2> &=& 4\pi \int \rho(r) r^4 dr= \Gamma({10\over3})(2a)^{-4/3},     
\end{eqnarray}
where $Q$ means the electromagnetic charge of the meson and $\Gamma$ designates 
the gamma function.  Since the relative distance is not the real charge radius
of the pion, the density should be replaced 
to $\rho(r)=Q_{\bar{u}}|\Psi_{\bar{u}}(r_{\bar{u}})|^2
+Q_d|\Psi_d(r_d)|^2$ with $r_u=m_dr/(m_u+m_d)$ and
$r_d=m_ur/(m_u+m_d)$.  The charge radii for the pion and kaons are calculated
as
\begin{eqnarray}
<r_{\pi^-}^2> &=& {1\over3} \Gamma({10\over3}) 
  {2 m_d^2+m_u^2 \over{(m_u+m_d)^2}}({9 \over{32 E_r^u k}})^{2/3}, \nonumber\\
<r_{K^-}^2> &=& {1\over3} \Gamma({10\over3}) 
  {2 m_s^2+m_u^2 \over{(m_u+m_s)^2}}({9 \over{32 E_r^u k}})^{2/3}, \nonumber\\
<r_{K^0}^2> &=& - {1\over3} \Gamma({10\over3}) 
  {m_s-m_d \over{m_s+m_d}}({9 \over{32 E_r^d k}})^{2/3}, \label{radii}
\end{eqnarray}
where the relative energy is calculated as $E_r^1=(M^2-m_1^2-m_2^2)/2(m_1+m_2)$ 
as mentioned previously from the energy $E'_1=(M^2-m_1^2-m_2^2)/2m_2$ 
in the rest frame of $m_2$, and $E'_1$ is calculated from the Lorentz 
invariance of the momentum squared
$(p_1+p_2)^2_{\text{c.m.}}=(p'_1+p'_2)^2_{\text{rest frame of $m_2$}}$ and 
the energy calculated below in the center of mass frame.  The index 1 of the
two particles is assigned to the lighter one through out this letter.  
Since the number of the unknown 
parameters is more than that of the equations, more physical processes are 
needed to calculate.  So the decay rates for 
the above meson are calculated in the following.

From the relation $dN=\rho v d\sigma$, where $N$ is the number of transition
particles and $v$ is the relative velocity between the two initial particles, 
the differential decay rate can be defined by
\begin{equation}
d\Gamma = {S|\Psi(0)|^2 \over{2E_1 2E_2}}|{\cal M}|^2 (2\pi)^4 \delta^4(p_1+
p_2-k_1-k_2){d^3k_1 \over{(2\pi)^3 2\omega_1}}
{d^3k_2 \over{(2\pi)^3 2\omega_2}},
\end{equation}
where $S$ is a symmetric factor for the initial particles and 
$|\Psi(0)|^2=4E_r k/3\pi$ from the Eq. (\ref{wave}).  The invariant 
amplitude for the pion decay process $d(p_2)+\bar{u}(p_1) \rightarrow\mu^-(k_2)+
\bar{\nu}(k_1)$ is calculated as
\begin{equation}
|{\cal M}|^2=32 G_F^2 \cos^2 \theta H_{\mu \nu}L^{\mu \nu},
\end{equation}
where the hadronic tensor means that $H_{\mu \nu}={1\over2}Tr[\bar{u}(p_1)
\gamma_{\mu}{1\over2}
(1-\gamma_5)u(p_2)\bar{u}(p_2)\gamma_{\nu}{1\over2}(1-\gamma_5)u(p_1)$ and
the leptonic tensor is similar to that for the momentum $-k_1$ and
$k_2$.  The integration of the product of the tensors over the whole solid angle
is calculated as
\begin{eqnarray}
\int H_{\mu \nu}L^{\mu \nu} d\Omega&=& \int 4 p_1 \cdot k_1 p_2 \cdot k_2 
d\Omega \nonumber\\
&=& 64 \pi E_1 E_2(\omega_1\omega_2+{p^2k^2\over{3 E_1 E_2}}),
\end{eqnarray}
where the energies and momenta are calculated from the relation $q^2=M^2=(p_1+
p_2)^2=(k_1+k_2)^2$ in the center of mass frame as
\begin{eqnarray}
\mbox{\boldmath{$p$}}^2&=&{1\over{4M^2}} \{(M^2-m_1^2-m_2^2)^2-4m_1^2m_2^2 \},
\nonumber\\
E_1&=&{1\over{2M}} (M^2+m_1^2-m_2^2),\nonumber\\
E_2&=&{1\over{2M}} (M^2+m_2^2-m_1^2),\nonumber\\
\mbox{\boldmath{$k$}}^2&=&{1\over{4M^2}} (M^2-m_{\mu}^2)^2, \nonumber\\
\omega_1&=&{1\over{2M}} (M^2-m_{\mu}^2),\nonumber\\
\omega_2&=&{1\over{2M}} (M^2+m_{\mu}^2),
\end{eqnarray}
where $M$ is the meson mass which is the bound system of $m_1$ and $m_2$.
From the above knowledge, the total decay rate is calculated as
\begin{equation}
\Gamma={G_F^2 \over{8 \pi}} |V_{12}|^2 m_{\mu}^2M(1-{m_{\mu}^2 
\over{M^2}})^2
[{32SME_r^1 k \over{3\pi m_{\mu}^2}} 
\{1+{m_{\mu}^2 \over{M^2}}+{(1-{m_{\mu}^2\over{M^2}})
((1-{m_1^2 \over{M^2}}-{m_2^2 \over{M^2}})^2-{4m_1^2m_2^2 \over{M^4}}) 
\over{3(1-({m_1^2 \over{M^2}}-{m_2^2 \over{M^2}})^2)}}  \}],\label{decay}
\end{equation}
where the terms in the square bracket correspond to the pion decay 
constant $f_{\pi}^2$
in the usual calculation for $\pi^- \rightarrow \mu^-+\bar{\nu}$ and the kaon
decay constant $f_K^2$ for $K^- \rightarrow \mu^-+\bar{\nu}$.
From the comparison between the decay and the charge radius of the pion in Refs. 
\cite{Weiss,Kokkedee}, the space wave function $|\Psi(0)|^2$ in the radius 
is 19 times larger than that in the decay.  The reason is conjectured 
that the symmetric factor is not properly considered and the form factor of 
the pion is a dipole type in the reference.  As the numerical calculations are 
shown below, the space wave function in the charge radii 
agrees to that in the decay rate strikingly, if the symmetric factor $S$ is 
assumed to be $1/(4 \times 9)$.  The $1/4$ means that the pseudoscalar meson is
in the spin signet state, which may have been considered in Refs. 
\cite{Weiss,Kokkedee} already.  However the color factor $1/9$, which means
that the decay takes place in the color singlet state and does not occur during
the interactions of the exchange of 8 gluons, must have been
ignored.

For the typical current quark masses $m_u=10$ MeV, $m_d=28.61$ MeV and 
$m_s=160$ MeV, the numerical values are calculated and compared with the 
experimental values \cite{Amendolia,Amendol,Molzon,European} in the second
column of Table \ref{table}, 
using the value $k=34655.5~ \text{MeV}^2$ from the relation just above 
Eq. (\ref{wave}) 
with the estimated value $\alpha_s=13.35$ in Ref. \cite{Ko}.  The reason for
using the same value $k$ through out all the physical quantities in the table
is that the mass of the
strange quark is known to be so heavy that it can not satisfy the condition
$4 m_s^2<q^2$.  Instead of the decay rate, the decay constants are regarded as 
the experimental values for simplicity.  The variation of the factor in the 
brace bracket in Eq. (\ref{decay}) is much small as the variation of the quark
mass, actually it is less than 0.1 around its values 1.70 for the pion decay 
and 1.30 for the kaon decay. Hence the u- and d-quark mass ratio is extracted 
as 0.35 from
the comparison of Eqs. (\ref{radii}) and (\ref{decay}) for the common $E_r^1k$ 
in the case of the pion, but the u- and s-quark mass ratio does not give 
an acceptable value in case of the kaon because
of the small value of the kaon decay constant. If the calculated value of the
kaon decay constant is replaced to the experimental value, the the u- and 
s-quark mass ratio is calculated as 0.065 and the CKM matrix element 
$|V_{us}|$ is estimated as 0.101.  This matrix element is out of the
range of the present standard values 0.217-0.224 \cite{European,Review}.  
However there may be 
sufficient room to account for this discrepancy, if the $K_{e3}$ decay
$K^+ \rightarrow \pi^0+e^++\nu$ is reconsidered carefully with an elegant 
quark model and the CKM matrix
is chosen as the Kobayashi and Maskawa's original parameterization.
This numerical calculation gives a new relation 
$f_{\pi}/f_K \approx M_{\pi}/M_K$.

This comparison of the equations
can be applied to other models, such as the wavefunction $\Psi(r)=\sqrt{a^3
/{\pi}}e^{-ar}$ for the Coulomb potential $V_{\text{eff}}=-{\alpha /r}$
and $\Psi(r)=({2a /{\pi}})^{3/4}e^{-ar^2}$ for the harmonic oscillator
potential $V_{\text{eff}}=br^2$.  If the above quark masses are adjusted to 
the charge radii and taken as the inputs, the decay constants of the pion and 
kaon are 1.55 and
2.77 MeV for the coulomb potential and 274.2 and 583.9 MeV for the harmonic
oscillator potential, respectively.  Since the exponential function of the
wavefunction reflects the asymptotic behavior of the potential, only the linear 
potential gives 
consistent results for the two different physical quantities, that is, charge
radii and decay rates.

Finally other current quark masses such as $m_u=1.5$ MeV \cite{European}, 
$m_d=4.29$ MeV, and $m_s=14.77$ MeV give also interesting results as shown 
in the third column of table \ref{table} for other suitable value $k=5057.66$ 
MeV$^2$ which means $\alpha_s=34.0$ in the point of view of Ref. \cite{Ko} as 
used previously.  These quark masses are adjusted to the new relation 
$f_{\pi}/f_K = M_{\pi}/M_K$.
Since the quark mass ratios are reproduced similar to that of the inputs,
it is more reliable to adjust the quark masses to the experimental values for
the charge radii rather than the decay constants.  If the quark masses are 
adjusted to the decay constants and 
taken as inputs, then the estimated values for the charge radii of kaons are 
too large to accept them with ignoring the experimental values.  It is also 
interesting to fit the quark masses to the experimental values for different 
values $k$ for the pion
and kaons, respectively, but it is still impossible to fit the experimental 
values to the charge radii and the decay constant of the kaons simultaneously
with the same current quark masses due to the small kaon decay constant.  This
is the principal reason that there are four parameter, that is, 
$m_u,~ m_d,~ m_s$, and $k$, and five physical quantities, namely, 
$<r_{\pi^-}^2>,~ <r_{K^-}^2>,~ <r_{K^0}^2>,~ f_{\pi}$, and $f_K$, but it is 
difficult to fix the parameters with those physical quantities.

\begin{table}
\begin{center}
\begin{tabular}{|c|c|c|c|}
physical quantities & calculated values I 
& calculated values II & experimental values  \\
\hline\hline
$<r_{\pi^-}^2>$   & 0.438 fm$^2$ & 
0.433 fm$^2$ & 0.43 $\pm$ 0.016 fm$^2$  \\
$<r_{K^-}^2>$   & 0.347 fm$^2$ & 
0.227 fm$^2$ & 0.34 $\pm$ 0.05 fm$^2$   \\
$<r_{K^0}^2>$   & - 0.145 fm$^2$ &
- 0.083 fm$^2$ & - 0.054 $\pm$ 0.026 fm$^2$   \\
$f_{\pi}$      &  129.38 MeV & 
131.15 MeV & 130.7 $\pm$ 0.1 $\pm$ 0.36 MeV   \\
$f_K$        &  347.52 MeV & 
463.89 MeV & 159.8 $\pm$ 1.4 $\pm$ 0.44 MeV  \\ 
\end{tabular}
\end{center}

\caption{The charge radius and the decay constant of the pion agree 
excellently to the experimental values, but those for the kaons do not
account for the experimental values properly.  The quark masses are adjusted
to the radius of $K^-$ in the first calculation, and to the relation
$f_{\pi}/f_K = M_{\pi}/M_K$, that is, the calculated $f_K$ in the second 
calculation.  The experimental values are taken from Refs. [10-14].}
\label{table}
\end{table}

\end{document}